# Characterizing segregation in blast rock piles: a deep-learning approach leveraging aerial image analysis

Chen-Geng Liu; Si-Hong Liu*; Chao-Min Shen*; Yu-Peng Gao; Yu-Xuan Liu

*College of Water Conservancy and Hydropower Engineering, Hohai University, Nanjing 210098, China. E-mail address: chengengliu@hhu.edu.cn (C.-G. Liu); sihongliu@hhu.edu.cn(S.-H. Liu); chaomin.shen@hotmail.fr (C.-M. Shen); gaoyp@hhu.edu.cn (Y.-P. Gao); liuyx0427@hhu.edu.cn (Y.-X. Liu)*

** Corresponding Authors at: College of Water Conservancy and Hydropower Engineering, Hohai University, Nanjing 210098, China. E-mail address: sihongliu@hhu.edu.cn(S.-H. Liu); chaomin.shen@hotmail.fr (C.-M. Shen)*

**Abstract:** Blasted rock material serves a critical role in various engineering applications, yet the phenomenon of segregation—where particle sizes vary significantly along the gradient of a quarry pile—presents challenges for optimizing quarry material storage and handling. This study introduces an advanced image analysis methodology to characterize such segregation of rock fragments. The accurate delineation of rock fragment size distributions was achieved through the application of an enhanced Unet semantic segmentation model integrated with an expansion-based post-processing technique. The quarry slope was stratified into four vertical sections, with the size distribution of each section quantified via ellipsoid shape approximations. Our results disclose pronounced vertical segregation patterns, with finer particles concentrated in the upper slope regions and coarser particles in the lower. The techniques outlined in this study deliver a scalable method for assessing fragment size distribution, with the potential to better inform resource management and operational decisions in quarry management.

## 1. Introduction

Blasted rock material represents a significant construction component widely utilized in numerous engineering fields, such as road subgrade construction [1, 2], dike construction [3], concrete batching [4], and soil and water conservation [5], among others. In the mining sector, blasted rock also exists in abundance as waste rock [6, 7]. Chamber blasting [8] is a common rock blasting method, renowned for its simplicity, minimal mechanical equipment requirement, low temporary construction workload, and large one-time blasting volume, making it particularly suitable for steep terrains and rock fields with developed rock fissures and joints. The blasted rock, hauled by haulage trucks, is dumped along the mountainside, forming extensive rock piles (slopes) serving as temporary quarries for subsequent use in engineering construction. Blasted rock is composed of coarse and fine particles of a wide range of sizes, and their size distribution refers to the proportion of rock particles within various size ranges [9]. Specific projects possess distinct requirements for the size distribution of blast rock, necessitating precise control over the size distribution to meet the actual construction requirements. Firstly, the selection of suitable blasting parameters during the blasting stage is essential. Of equal significance is the thorough planning and blending of batches of rock material, especially when there exists notable segregation along the height direction of the rock pile within the quarry.

Rock dumped along a mountainside tends to exhibit a significant segregation phenomenon: larger (i.e., heavier with stronger momentum) blocks tend to settle at the foot of the pile, while finer particles remain closer to the deposition point and the crest of the pile, thereby creating distinctive spatial heterogeneity [10]. Therefore, during the batching process of rock usage, if such spatial heterogeneity on the rock slope is overlooked, the batched rock may be excessively coarse or fine, which would not meet the construction design indicators. For instance, when overly fine rock is used in the construction of the subgrade in road construction, it may result in inadequate strength and poor drainage of the subgrade [11], significantly undermining its stability and durability. Therefore, the proper characterization of segregation phenomena is vitally important for planning rock blending. Moreover, this uneven particle size distribution closely relates to key geo-mechanical properties, such as permeability [12] and shear strength [13], further emphasizing the importance of addressing segregation in engineering analysis.

Fragment size distribution is a crucial source of data for characterizing segregation phenomena. Various techniques have been developed for assessing the size distribution of rock fragments. Common methods include visual observation, sieve analysis, and image-based analysis [14-16]. The visual method involves inspecting the rock pile to make a subjective evaluation of the blasting quality, which could yield inaccurate results. Sieve analysis involves obtaining a representative sample from the rock formation under investigation and subjecting it to sequential sieves of varying mesh sizes. The size distribution of the rock particles is then determined by quantifying the mass of the rock material retained on each sieve. Although renowned for reliability and precision, this method requires considerable time and financial resources and may be impractical when the sampled distribution does not accurately represent the aggregate composition of the entire rock pile. Image analysis presents a feasible strategy for particle size distribution analysis, and has been widely applied in multiple research fields, such as aggregate fragmentation [17, 18] and clay fracture analysis [19]. This approach typically entails automatic extraction of particle information, such as shape, area, and axis lengths, and then utilizes approximation methods to calculate the three-dimensional volume from two-dimensional fragment images before engaging in fragment size distribution statistics.

### *1.1 Related work*

In recent years, a large amount of research work has employed image analysis and numerical simulation to study the segregation phenomena of particles, including blasted rock. Zhang et al. [20] utilized aerial image analysis to investigate particle size distribution within ore piles, critically informing our understanding of fluid flow behaviours through packed ore/rock beds. Their analysis revealed a substantial variance in particle size, along with a spatial segregation of fine and coarse particles within the leach pad. Qiu et al. [21] utilized a methodological approach involving threshold segmentation in image analysis to scrutinize particle size distribution of a waste rock pile, delineate PSD curves for distinct sections along the slope, and quantitative assess segregation degrees. Dai et al. [22] conducted a series of physical model experiments and DEM simulations to inspect the particle sorting feature in scree slopes. They proposed relative particle size and shape factors to explain the sorting feature delineated by a novel segregation index. Ye et al. [23] carried out a series of laboratory-scale experiments to characterize the size segregation of stockpiles, explored the degree of segregation under different drop heights and particle top sizes, and emphasized the feasibility of directly scaling laboratory-scale to industry-scale. The characterization of segregation phenomena can furnish valuable information for operators in industrial stockpiles. Qiu et al. [10] advanced the field of waste rock disposal by introducing a new calibration technique for DEM simulations and Their calibrated DEM models successfully replicated observed segregation in waste rock piles. With regard to real large-scale rock slopes, image analysis (primarily image segmentation) exhibits a significant advantage in characterizing particle segregation, due to its ability to process vast data sets rapidly and provide detailed spatial information on material composition. However, most of the image segmentation methods used in the aforementioned research are simple threshold segmentation, which suffers from poor segmentation accuracy and is easily affected by environmental lighting, leading to error in size analysis. As the acquisition of accurate fragment size distribution depends on precise image segmentation, it is vital to explore better image segmentation methods for rock images.

Research on particle segmentation has shown a notable growth trend in recent years. Compared to traditional image segmentation algorithms such as watershed [14, 24-26] , image segmentation algorithms based on deep convolutional networks have a distinct advantage in accuracy [27-30]. Specifically in the context of rock image segmentation, Yang et al. [31] proposed an enhanced Unet model for rock pile segmentation. The model, built upon the original Unet, incorporates depthwise separable convolution and batch normalization to reduce model size and accelerate model convergence during training, achieving an average segmentation accuracy increase of 1.53% compared with the original Unet. Guo et al. [32] integrated Unet, ResNet, and ASPP to put forward an FRRSnet+ network for the segmentation of blasted rock fragments. This method consists of two models, one of which is responsible for complementing the rock edge. The model’s segmentation accuracy displayed various degrees of improvement compared to Unet and Segnet. Shrivastava et al. [33] utilized Mask R-CNN to segment particles of mine overburden dump, accurately detecting an average of 73.58% of particles in each image with a mean percentage error of 0.39%. Semantic segmentation algorithms, such as Unet, exhibit robust performance and superior inferencing speed compared to instance segmentation algorithms like Mask R-CNN; as such, they are frequently used in particle segmentation tasks. One limitation of semantic segmentation algorithms is the difficulty in distinguishing boundaries when substantial overlap exists between particles; it may recognize several regions as a single region, resulting in under-segmentation. This issue can be rectified by separately detecting edge pixels.

### *1.2 Contributions of this work*

In this study, we acquire aerial images of rock fragments on the slope of the quarry using drones to characterize the particle size distribution and segregation phenomena of quarry rock particles. Specifically, we propose an enhanced Unet model integrated with a region expansion-based post-processing method to achieve rock particle segmentation. The entire rock slope is divided into several regions, and the particle size distribution and characteristic particle sizes of each region are calculated, including $d_{10}$, $d_{50}$, and $d_{90}$. Finally, the characteristic particle sizes of the original rock material ($d'_{10}$, $d'_{50}$, $d'_{90}$) are calculated, and the segregation phenomena are characterized using relative characteristic diameters ($d_{10}/d'_{10}$, $d_{50}/d'_{50}$, and $d_{90}/d'_{90}$). Characterization of segregation phenomena can provide guidance for the sourcing and transporting of materials during construction.

## 2 Methodology

### *2.1 Overall framework*

To characterize rock segregation phenomena using aerial images of the rock, it is necessary to obtain the grain size distribution within the images. Using an improved Unet semantic segmentation model, we extract the body and boundary of the particles. Subsequently, a post-processing method based on region expansion is used to acquire each complete particle. Upon calculating the volume of the particles using equivalent ellipsoids, we then statistically analyze the grain size distribution at different height partitions. Finally, we characterize the segregation phenomena using indicators like relative characteristic diameters. Fig. 1 depicts the schematic of the overall proposed framework.

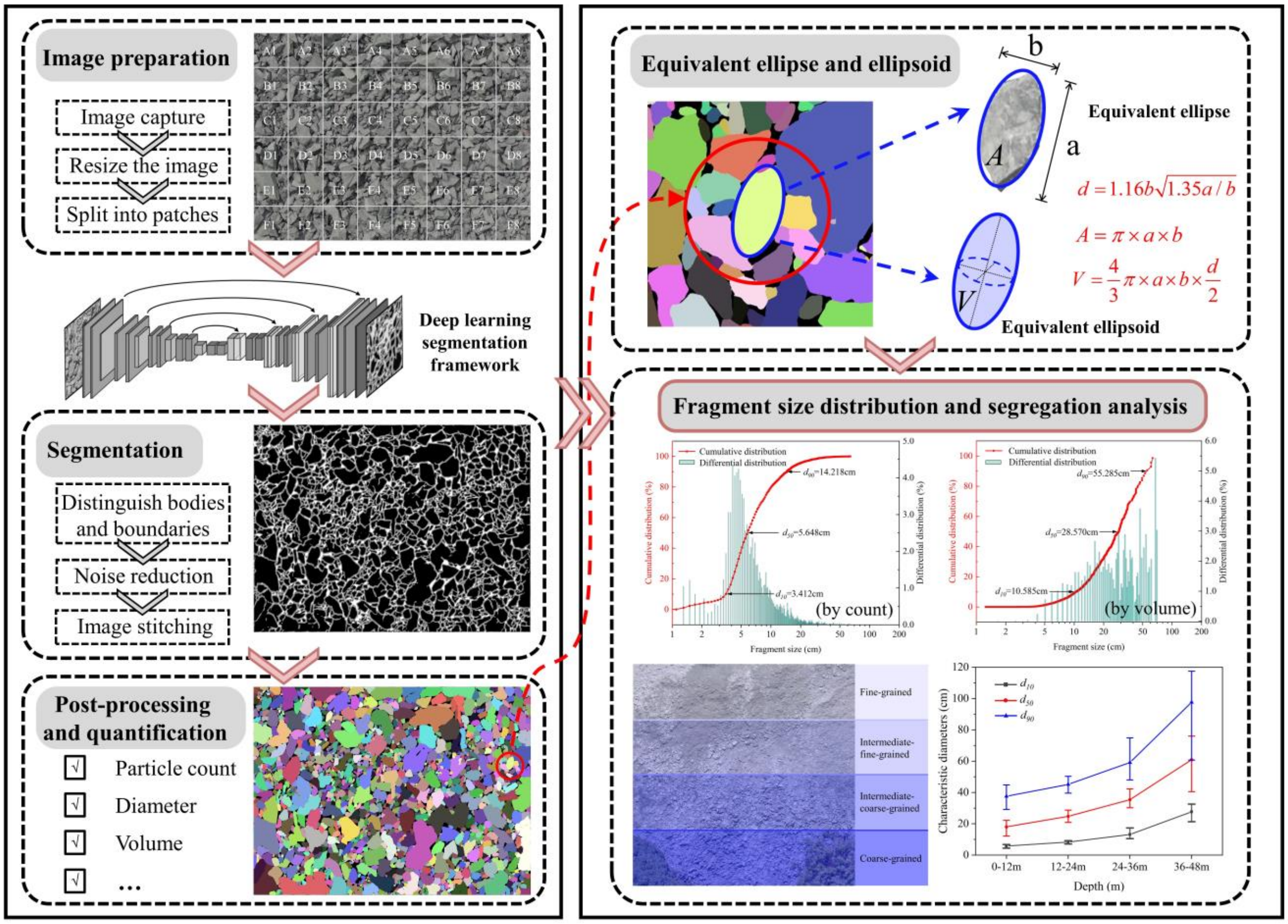

***Fig. 1.*** *Schematic of overall framework.*

### *2.2 Image acquisition and processing*

The case study site is located in a chamber blast quarry in Huangshan City, eastern China, as shown in Fig. 2 (a). The slope gradient of the quarry is approximately 38 degrees, and the distance from the top of the slope to the bottom measures around 48 meters. Photographs were taken using a drone (with specifications detailed in Table 1) that flew at a constant altitude above the slope face, ensuring that the camera lens was vertically aligned with the slope. Square aerial survey markers were employed to calibrate the scale between actual size and pixel dimensions. Fig. 2 (b) illustrates the schematic of the drone capturing images of the rock slope. Fig. 2 (c) displays a selection of the sample photographs captured.

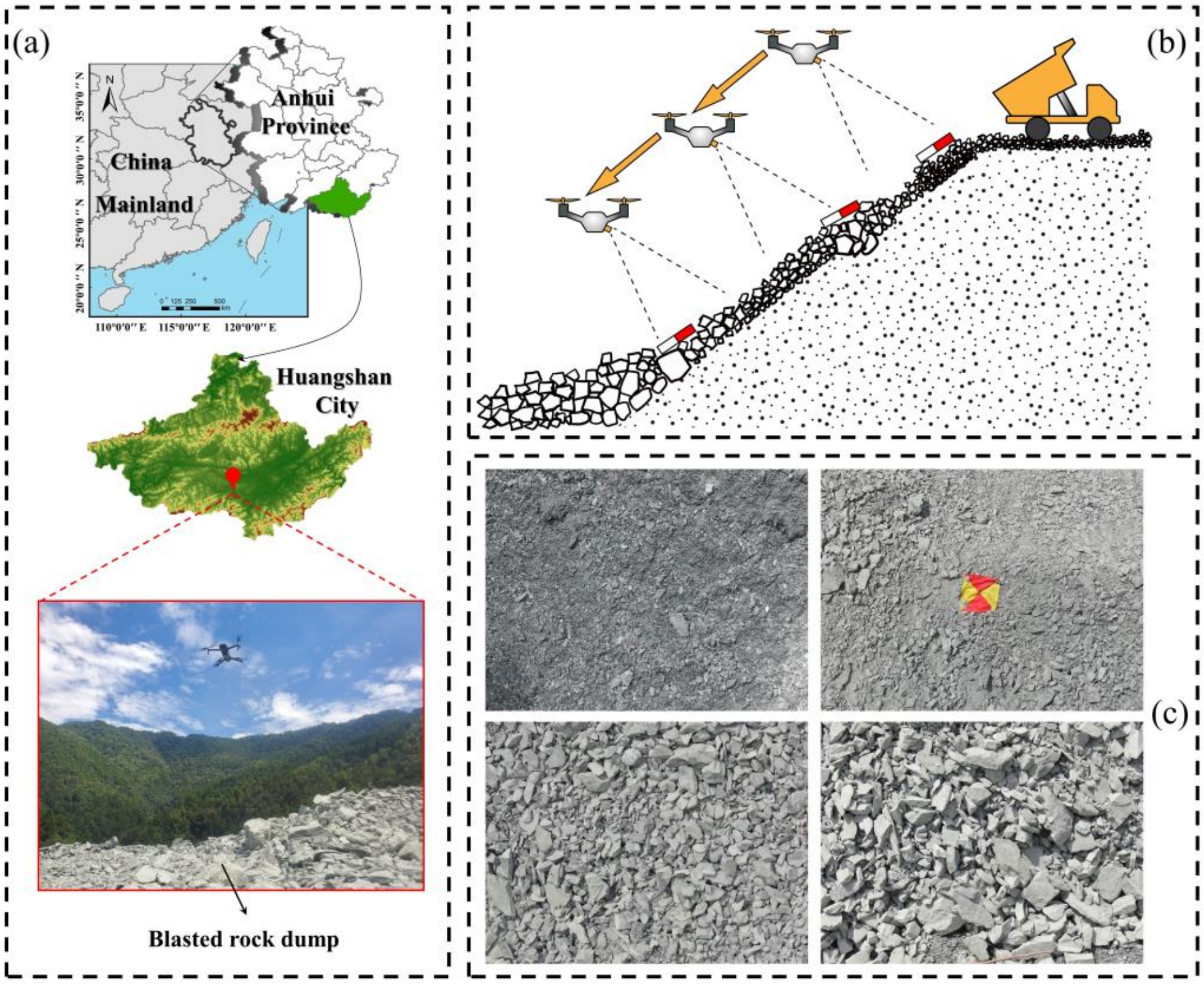

*__Fig. 2.__ Image acquisition: (a) the site of blasted rock dump; (b) diagram of image acquisition process; (c) sample images.*

***Table 1*** *Specifications of the UAV system and on-board camera*

| UAV system specifications | | On-board camera parameters | |
|---|---|---|---|
| UAV model | DJI Mavic 2 Zoom | Sensor | 1/2.3 inch CMOS, 12 MP |
| Weight | 905 g | Lens | FOV 83° (24mm); 48° (48mm) |
| Max flight time | 31 minutes | Frequency | 2.4 GHz |
| Satellite positioning | GPS/GLONASS/BeiDou | Photograph ISO range | 100-3200 |
| Gimbal stablization | 3-axis (pitch, roll, yaw) | Image size | 4000 × 3000 |

The captured images, originally sized at 4000 × 3000 pixels, were rescaled to 4096 × 3072 pixels using bilinear interpolation, and four images were selected to create the dataset for training the semantic segmentation model. These four images were cropped using a sliding window of 512 pixels with each adjacent window overlapping by 50%, resulting in a total of 660 image patches. A dual-labeling strategy was applied to semantically annotate the particles within these patches: the main body of particles was labeled with a 'body' tag, and the edges were labeled with a 'boundary' tag. The annotated patches were then divided into a training set and a validation set at a ratio of 8:2.

### *2.3 Particle segmentation*

#### *2.3.1 Proposed semantic segmentation network*

Unet [34], a widely recognized semantic segmentation network, features a symmetric U-shaped architecture consisting of a contracting path to capture the global context and an expansive path for accurate localization. It incorporates skip connections between the two paths to transfer context information to higher resolution layers. In this study, we utilized Unet with a VGG16 encoder [35] as the baseline framework. The downsampling block is repeated by two or three 3 × 3 cconvolutions (activated by ReLU functions) and one 2 × 2 max-pooling operation. Thus, the image is halved in size after each block, compensated with a doubled number of feature map channels. In the decoder part, repeated blocks include an upsampling (bilinear method), a concatenation with the corresponding feature map from the encoder, and two repeated 3 × 3 convolutions (each followd by a ReLU function) to fuse and reconstruct feature maps from both local details and global context. Ultimately, a 1 × 1 convolution, set to the number of classes, is employed to output class-wise classification results for each pixel.

In this study, Unet was employed as the foundational model, with three significant modifications conducted on the decoder. Fig. 3 illustrates the architecture of the enhanced Unet model. Firstly, we replaced the bilinear upsampling with the lightweight CARAFE upsampling operator. CARAFE boasts unique content-aware capabilities that enable precise recovery of image details while mitigating the loss of information about small targets often seen in the downsampling process, without necessitating the introduction of additional learning parameters [36]. Illustrated in Fig. 4, CARAFE comprises two primary modules: the kernel prediction module and the content-aware reassembly module. The objective is to convert the original feature map X (C × H × W) into the target feature map X′ (C × σH × σW). The positions $l' = (i', j')$ on X′ all correspond to the positions $l = (i, j)$ on X, where $i = \lfloor i'/\sigma \rfloor$ and $\lfloor j'/\sigma \rfloor$. The up-sampling kernel prediction module mainly predicts the upsampling kernel $w_{l'}$ at each positions $l'$ of the target feature map through a three-step operation of Channel Compressor, Content Encoder, and Kernel Normalizer. $N(X_l, k)$ denotes the $k \times k$ adjacent region of the feature map X centered at $l$ , as shown in Equation (1). The kernel prediction module first compresses the input feature map channels to reduce the subsequent computational load and then converts the number of channels to $\sigma^2 \times k_{up}{}^2$ by convolutional layers with a convolutional kernel size of $k_{encoder} \times k_{encoder}$, where σ is the upsampling ratio. Finally, the channel dimension is expanded in the spatial dimension and normalized with the sigmoid function to form an upsampling kernel of size $\sigma H \times \sigma W \times k_{up}{}^2$. The content-aware reassembly process is shown in Equation (2), where $r = \lfloor k_{up} / 2 \rfloor$. For the target location $l'$, the content-aware ressembly module first finds its corresponding region $N(X_l, k_{up})$ on the input feature map, and uses this region to do the dot product with the corresponding location of the up-sampling kernel in order to pay better attention to the information from the correlation points in the localized region, and generates a feature map with better semantic features.

$$w_{l'} = \Psi(N(X_l, k_{encoder})) \tag{1}$$

$$X'_{l'} = \sum_{n=-r}^{r} \sum_{m=-r}^{r} w_{l'(n,m)} \cdot X_{(i+n,\, j+m)} \tag{2}$$

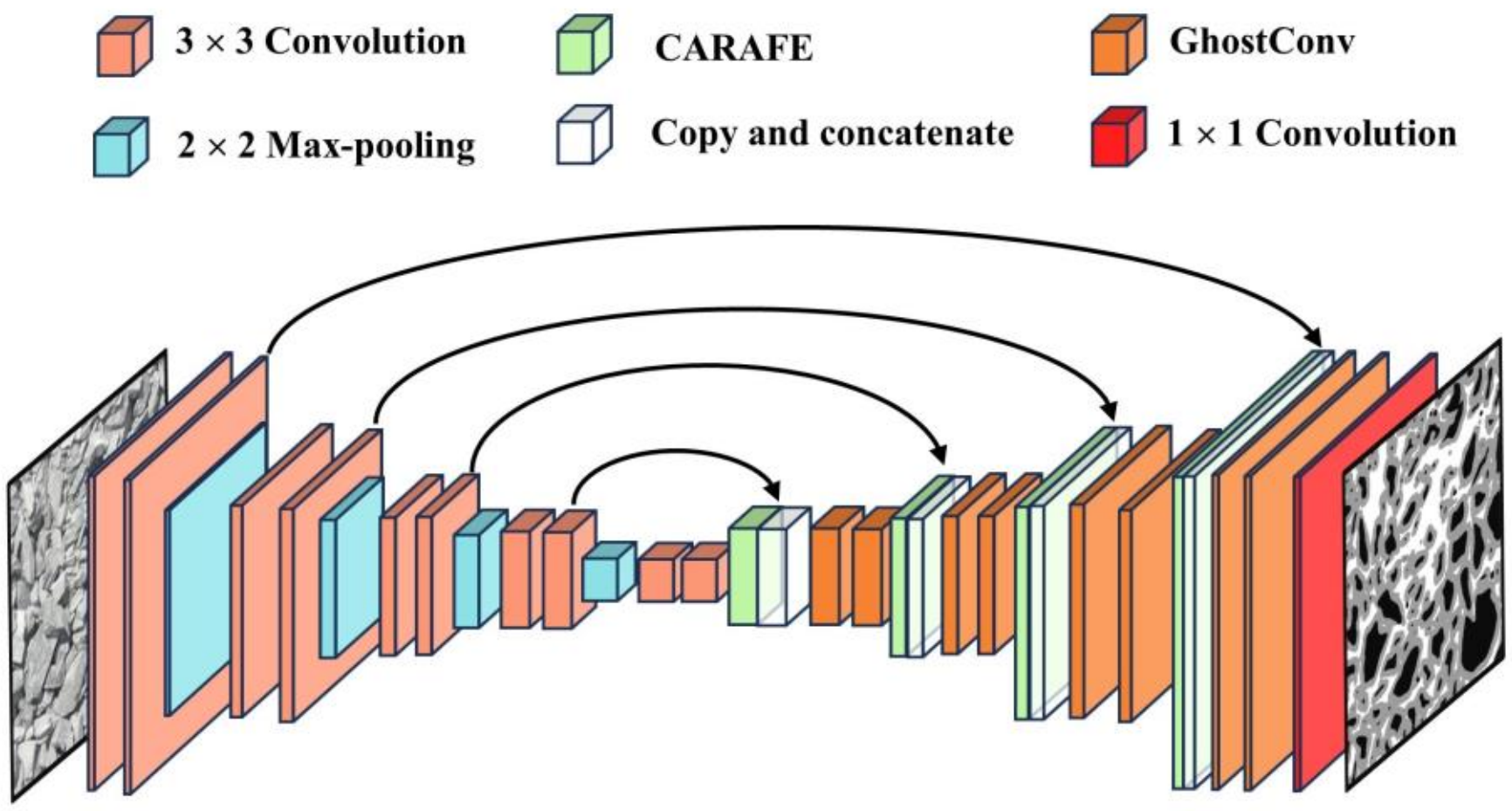

***Fig. 3.*** *Architecture of the improved Unet model.*

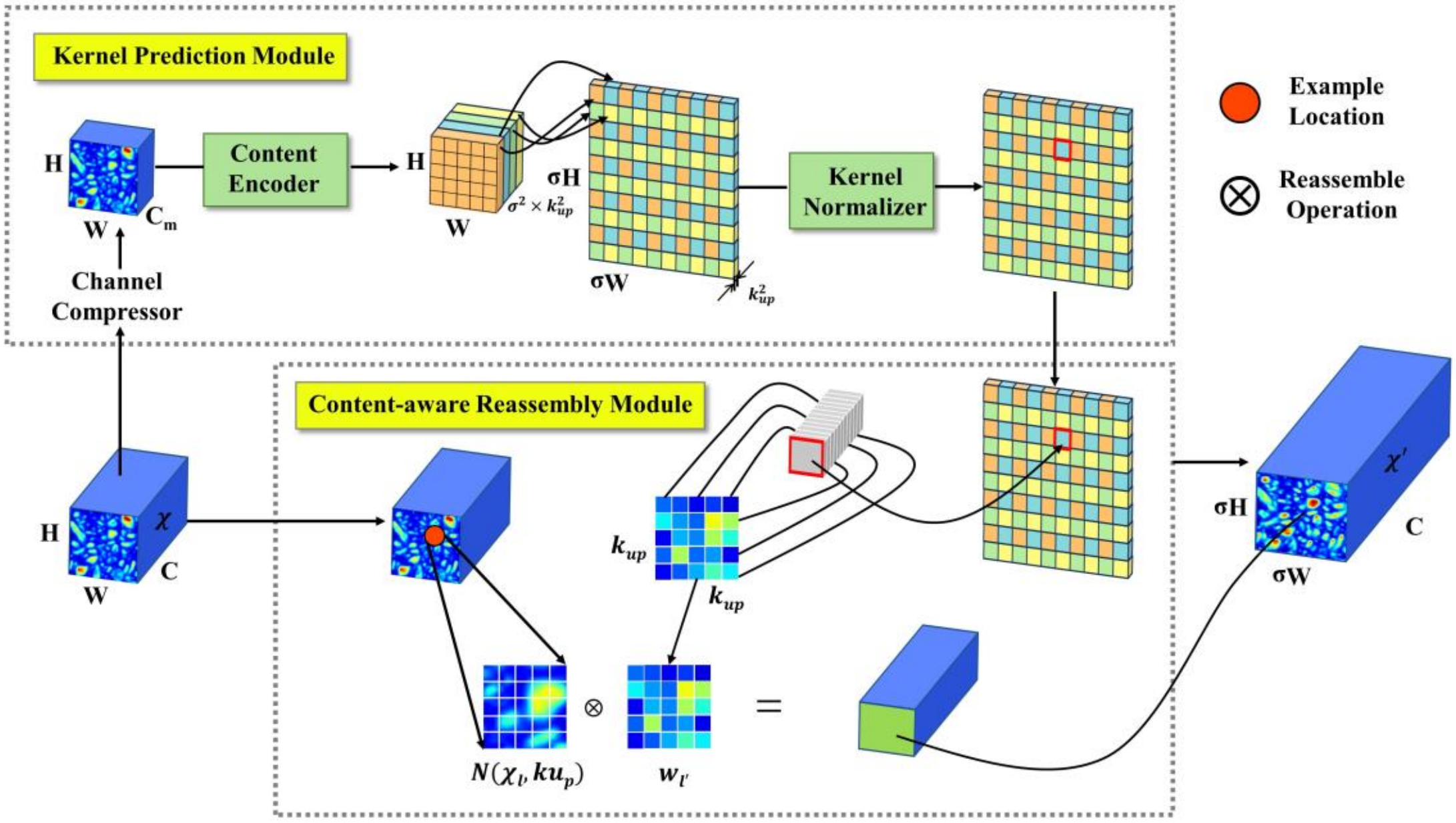

***Fig. 4.*** *Structure diagram of CARAFE module.*

The second modification addresses the issue that the convolutional operations in mainstream CNNs at various scales tend to produce a multitude of redundant feature maps. To tackle this, we replaced the two consecutive convolutional layers in the decoder stage with GhostConv [37] layers. These layers aim to generate a more efficient number of feature maps using fewer parameters and less costly operations. The structure of GhostConv module is shown in Fig. 5. Initially, input features undergo a channel reduction via a 1 × 1 convolution. This primary convolution step diminishes the dimensionality of the input features. Subsequently, the residuals are employed to construct the feature maps. The upper branch maintains the original information via direct mapping, while the lower branch deploys a 3 × 3 depth-separable convolution, grouping features based on their channels. Ultimately, the outputs from both branches are concatenated to finalize the feature extraction. By substituting the traditional convolutional layers in Unet's decoder with the GhostConv layers, the model becomes more lightweight and its detection accuracy can potentially be improved to a certain extent.

Furthermore, to enhance the feature extraction capability of our model, we incorporate an attention mechanism into the improved lightweight model. Specifically, we utilize the Efficient Channel Attention (ECA) [38] mechanism and integrate it into the GhostConv module. ECA is an exceptionally efficient and lightweight channel attention mechanism, consisting of three primary steps, as depicted in Fig. 6 (a). Initially, a global average pooling operation is applied to the input feature maps, effectively compressing the spatial dimensions W × H to 1 × 1 while preserving channel-wise information. Subsequently, a one-dimensional convolution with a kernel size of 3 is utilized to facilitate local cross-channel interaction and capture channel-wise dependencies. Following that, a sigmoid activation function computes the channel-wise attention weights. The final step involves multiplying the attention weights element-wise with the input feature maps, thereby enabling the network to selectively highlight pertinent information. In this study, ECA is integrated into each GhostConv module within the up-sampling stage. The modules incorporating this integration are referred to as Ghost-ECA, and the internal structure of Ghost-ECA is illustrated in Fig. 6 (b).

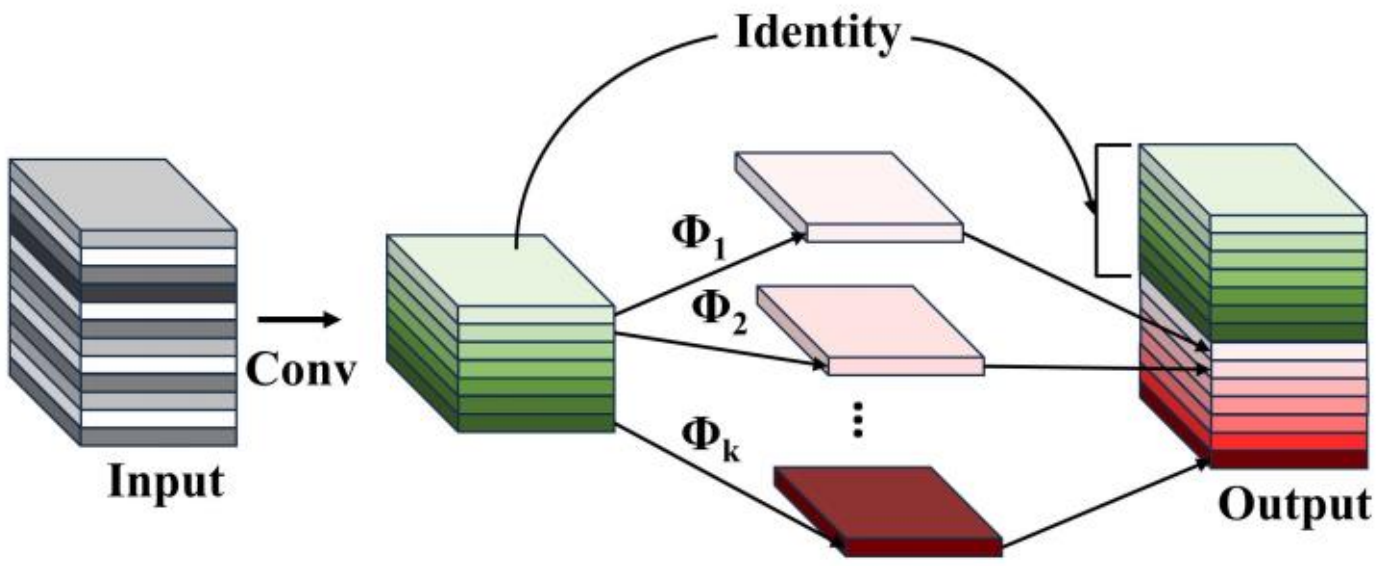

***Fig. 5.*** *Structure diagram of GhostConv module. Φ represents the cheap operation.*

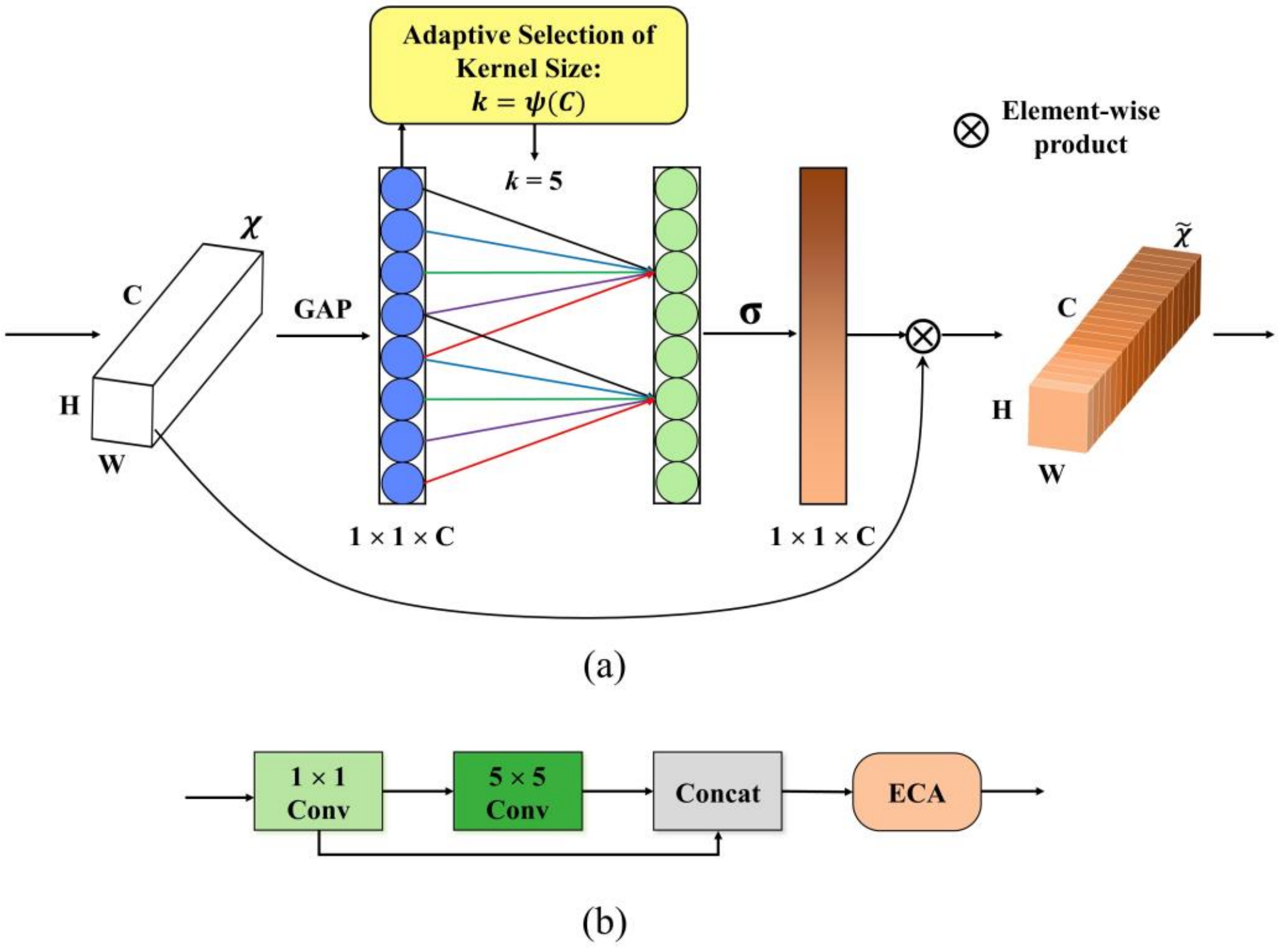

***Fig. 6.*** *(a) Diagram of ECA module. Given the aggregated features obtained by global average pooling (GAP), ECA generates channel weights by performing a fast 1D convolution of size k, where k is adaptively determined via a mapping of channel dimension C; (b) Internal structure of Ghost-ECA.*

*2.3.2 Image cropping and stitching*

With the trained semantic segmentation model, we can proceed to predict the segmentation of new images. Due to the large dimensions of the complete images that need to be predicted, directly processing full-size images poses challenges related to memory and computational resource limitations. Consequently, we employ a crop-and-stitch prediction method. During the model prediction phase, the full-size image (4096 × 3072) is divided into a series of 512 × 512 patches, with each adjacent pair of patches overlapping by 50%. After performing model predictions on these patches, the outer portions of each patch are discarded, retaining only the central 256 × 256 pixel region. Ultimately, these 256 × 256 pixel blocks are stitched together to form the segmented prediction results for the whole-size image. Fig. 7 illustrates the process described above. This method also addresses the common issue of prediction inaccuracies at the image edges (edge effects) by preserving the predicted results of the central regions of each image block, leveraging the high-quality predictions of these segments to enhance the accuracy of the overall stitched image prediction.

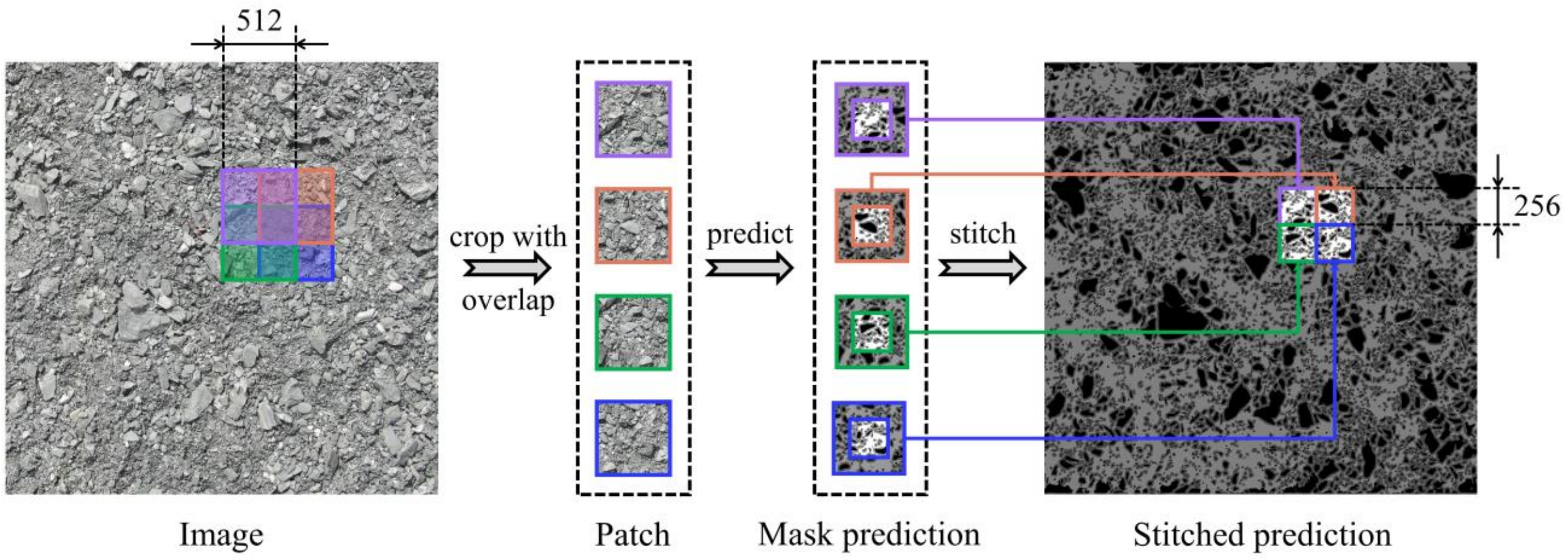

***Fig. 7.*** *Schematic diagram of image cropping and stitching for full-size image mask prediction.*

*2.3.3 Post-processing and shape estimation*

During the image annotation phase, the particles' bodies and boundaries were delineated. The primary function of the semantic segmentation model is to differentiate these two types of pixel regions. Subsequent to this classification, post-processing techniques are essential to merge the body and boundary segments to reconstruct each complete rock fragment. Additionally, this fusion facilitates further shape estimation and size information extraction.

We propose a straightforward region expansion strategy for the fusion of the body and boundary segments. This approach consists of three steps: (1) Designating each body region as a seed point for area expansion; (2) Setting a maximum expansion radius, which defines the farthest distance the region can expand outward from the edge of a seed point; (3) Establishing termination conditions for expansion, which include stopping the expansion when it extends beyond the boundary pixel range or encounters an expansion front originating from another seed point. Upon completing the area expansion operation, the fusion of the body and boundary segments is achieved, thus forming each complete rock fragment shape, as depicted in Fig. 8.

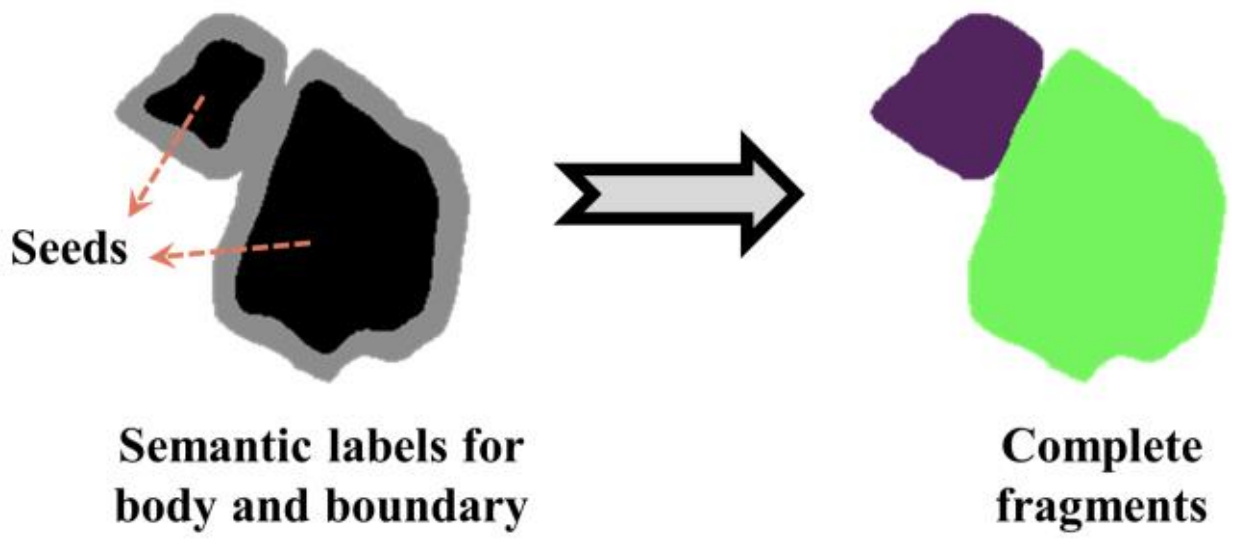

***Fig. 8.*** *Schematic diagram of obtaining complete fragments through post-processing.*

After segmenting individual fragments, automatic shape and size estimation of the fragments is conducted. Each particle contour is approximated by an equivalent ellipse, with the particle size denoted by the equivalent ellipse diameter, *d*. The diameter *d* is calculated based on the major axis, *a*, and minor axis, *b*, of the equivalent ellipse, as shown in Equation (3). The area, A, of the equivalent ellipse, which matches the area of the segmented particle, can also be expressed as $A = \pi \times a \times b$ accordingly. To obtain a volume-based size distribution, each fragment is considered as an ellipsoid, and its volume is calculated according to Equation (4).

$$d = 1.16b\sqrt{1.35a / b} \tag{3}$$

$$V = \frac{4}{3}\pi \times a \times b \times \frac{d}{2} \tag{4}$$

***2.4 Characterization of segregation***

After accomplishing particle segmentation and size calculation, it becomes feasible to compile statistics on particle size distribution, which facilitates the characterization of segregation phenomena. During the previous image acquisition phase, the entire quarry was proportionally divided into four vertical zones along the slope depth: S1, S2, S3, and S4, with each zone spanning approximately 12 meters along the slope. In each section, several non-overlapping images were captured, and the particles contained within these images are considered representative of the entire rock slope's particle composition. The particle size distribution aggregated from these images approximates the overall particle size distribution of the slope, referred to as the overall PSD curve. This curve is used to evaluate the segregation of rock fragments along the slope within each image. For the analysis of segregation, on the one hand, the particle size distribution in each section, including both number distribution and volume distribution, is statistically compiled and compared horizontally among sections. On the other hand, the relative characteristic diameters are calculated based on the characteristic particle sizes of each section compared to the overall characteristic particle sizes. These calculations are used to further characterize the segregation phenomena.

## 3. Experiments and results

### *3.1 Segmentation results*

#### *3.1.1 Model training*

The training environment is configured with PyTorch 1.11.0 and Python version 3.8.10, utilizing an RTX 3090 GPU with 24GB of memory. Detailed training parameters are listed in Table 2. A fixed random seed was used for each training session to ensure reproducibility. To enhance the model's robustness and generalization ability, data augmentation techniques were applied during training [39]. These techniques include random rotations, random flips, addition of noise, HSV (Hue, Saturation, Value) transformations, and Gaussian blur, as illustrated in Fig. 9. The utilized loss function encompasses both the cross-entropy loss and the Dice loss. The cross-entropy loss, which is pivotal in semantic segmentation, quantifies the discrepancy between the predicted probability distribution of each pixel class and the actual class label vector. The formula for cross-entropy loss is as follows:

$$L_{ce} = -\frac{1}{N}\sum_{i=1}^{N}\sum_{k=1}^{K} y_{i,k} \ln p_{i,k} \tag{5}$$

where $N$ represents the number of pixels in the image, $K$ represents the number classes in the image, $p_{i,k}$ is the probability of predicting the $i$th pixel as the $k$th class, and $y_{i,k}$ indicates whether the prediction aligns with the true class label for that pixel.

The Dice loss is primarily a measure of the accuracy of the distribution of each class, which can be expressed as follows:

$$L_{dice} = 1 - \frac{2\sum_{i=1}^{N}\sum_{k=1}^{K} g_i^k s_i^k}{\sum_{i=1}^{N}\sum_{k=1}^{K} g_i^k + \sum_{i=1}^{N}\sum_{k=1}^{K} s_i^k} \tag{6}$$

where $s_i^k$ , $g_i^k$ denote the prediction result and the true lable of the *k*th category for the *i*th pixel, respectively.

The total loss function can be expressed as:

$$\boldsymbol{L_{total} = L_{ce} + L_{dice}} \tag{7}$$

After completing 100 epochs of training, the loss curves have shown convergence. Fig. 10 illustrates the variation trends of the loss curves for five different models (Model 1, Model 2, Model 3, Model 4, and Model 5 respectively represent the Unet baseline model, the model with CARAFE, the model with GhostConv, the model integrating both CARAFE and GhostConv, and the model employing both CARAFE and Ghost-ECA).

***Table 2*** *Model training parameters*

| Parameters | Value |
|---|---|
| Innitial learning rate | $10^{-4}$ |
| Optimizer | Adam |
| Momentum | 0.9 |
| Batch size | 6 |
| Training epochs | 100 |

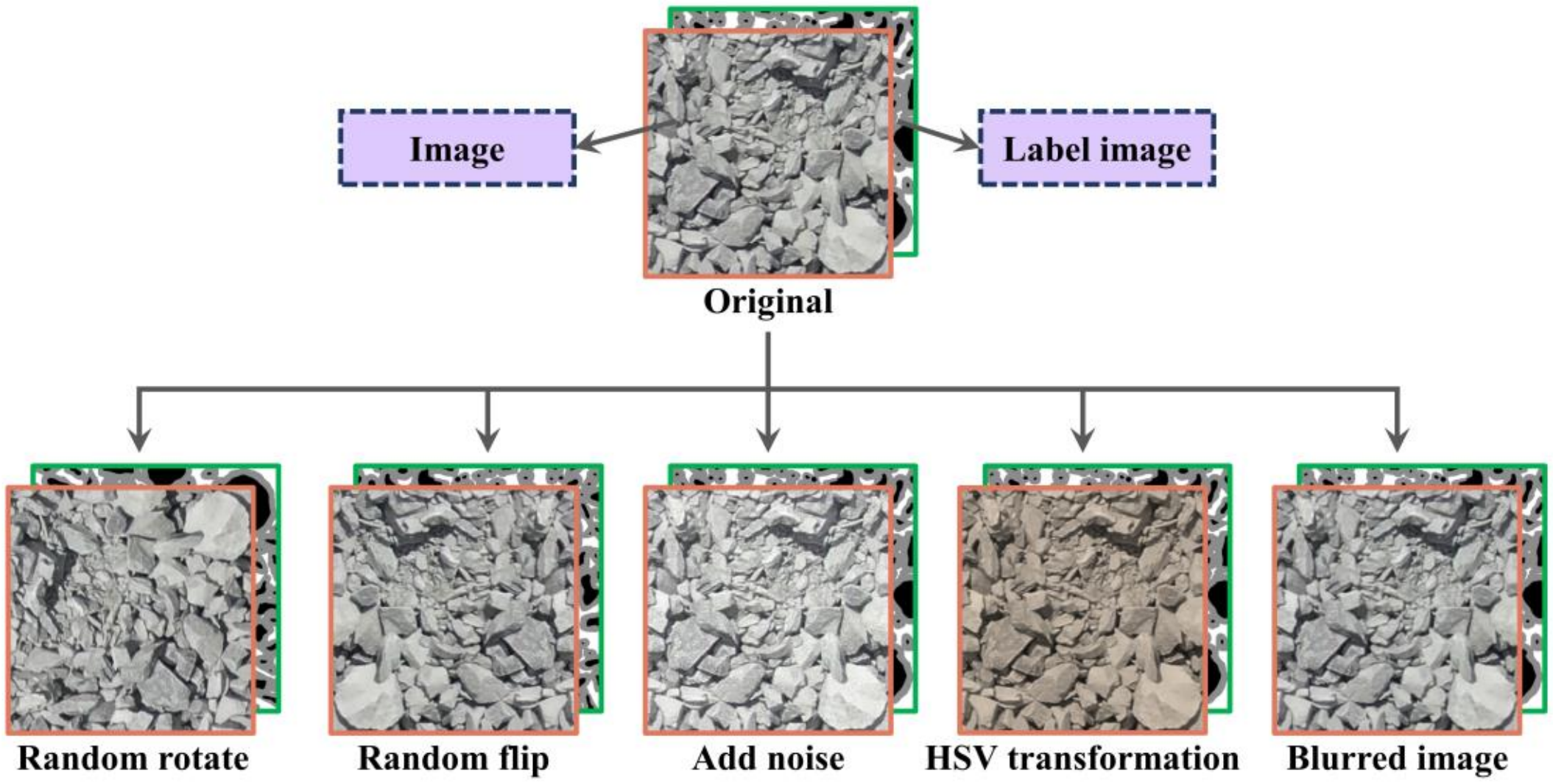

***Fig. 9.*** *Data augmentation tricks.*

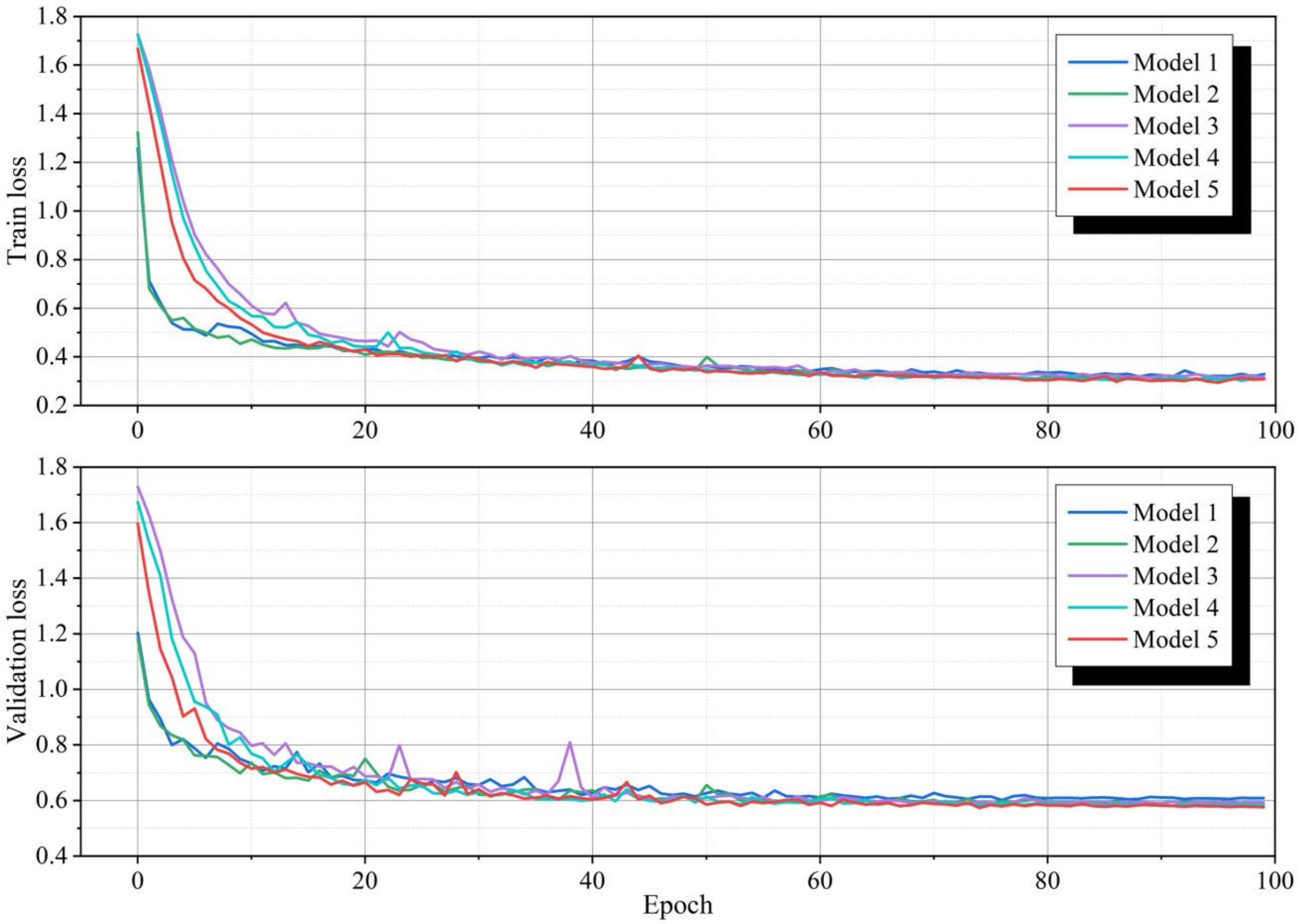

**Fig. 10.** *The loss curves.*

*3.1.2 Model evaluation*

Using the trained models, we performed predictions on test image patches to identify the body and boundary pixels of rock fragments. To evaluate the prediction accuracy of the proposed models and compare their performance, we selected several metrics such as Precision, Recall, F1, Mean Intersection over Union (mIoU), and Mean Pixel Accuracy (mPA). All these metrics are derived from the confusion matrix of predicted outcomes, which includes four components: true positive (TP), false positive (FP), true negative (TN), and false negative (FN). The formulas for these metrics are as follows:

$$Precision = \frac{TP}{TP + FP} \tag{8}$$

$$Recall = \frac{TP}{TP + FN} \tag{9}$$

$$F1 = 2 \times \frac{Precision \times Recall}{Precision + Recall} \tag{10}$$

$$IoU = \frac{TP}{TP + FP + FN} \tag{11}$$

$$mPA = \frac{TP + TN}{TP + TN + FP + FN} \tag{12}$$

IoU is the ratio of intersection and concatenation of predicted and true values, and F1 is the reconciled average of precision and recall. mIoU, mPA represents the mean of all classes IoU and pixel accuracy.

Table 3 showcases a quantitative comparison of Models 1 through 5. The improved Unet algorithm demonstrates superior performance compared to the baseline model, with mIoU and mPA reaching 72.18% and 83.46%, respectively, indicating an increase of 1.38% and 0.74%. Each enhancement module contributes to various extents. Notably, the improved model shows a significant boost in the accuracy of boundary recognition. The IoU metric for boundary recognition has improved by 1.96% over the baseline model, achieving 68.28%. In terms of model complexity, the updated model displays a reduced parameter count of 20.71M, representing a reduction of 16.8% from the 24.89M parameters of the baseline model.

To vividly illustrate the contribution of each module to the enhancement of model performance, Fig. 11 visualizes the feature maps of the various models, with the body and boundary features displayed separately. It is evident that the improvement strategies we proposed significantly strengthen the model's capability for extracting features of both the body and boundary, particularly for the boundary. The responsiveness of the edge features has been notably heightened, leading to more precise delineation of particle edges. This indicates that our enhancement strategies have effectively improved the model's ability to capture information regarding the shape of particle edges, which ensures greater accuracy in subsequent post-processing tasks.

***Table 3*** *Results of ablation experiments. The best results are shown in bold.*

| Model No. | CARAFE | Ghost-Conv | mIoU | mPA | Body | | | | Boundary | | | |
|---|---|---|---|---|---|---|---|---|---|---|---|---|
| | | | | | IoU | Precision | Recall | F1 | IoU | Precision | Recall | F1 |
| 1 | × | × | 70.80 | 82.72 | 82.37 | 89.46 | 91.23 | 90.34 | 66.32 | 81.30 | 78.25 | 79.75 |
| 2 | √ | × | 71.35 | 83.08 | 82.79 | 89.55 | 91.64 | 90.58 | 66.44 | 81.99 | 78.49 | 80.20 |
| 3 | × | √ | 71.52 | 83.46 | 82.88 | 89.79 | 91.50 | 90.64 | 66.88 | 82.24 | 78.16 | 80.15 |
| 4 | √ | √ | 71.77 | 83.34 | 83.26 | 89.57 | **92.20** | 90.87 | 67.44 | **82.72** | 78.50 | 80.55 |
| 5 | √ | √ (+ ECA) | **72.18** | **83.46** | **83.35** | **90.36** | 91.48 | **90.92** | **68.28** | 81.68 | **80.63** | **81.15** |

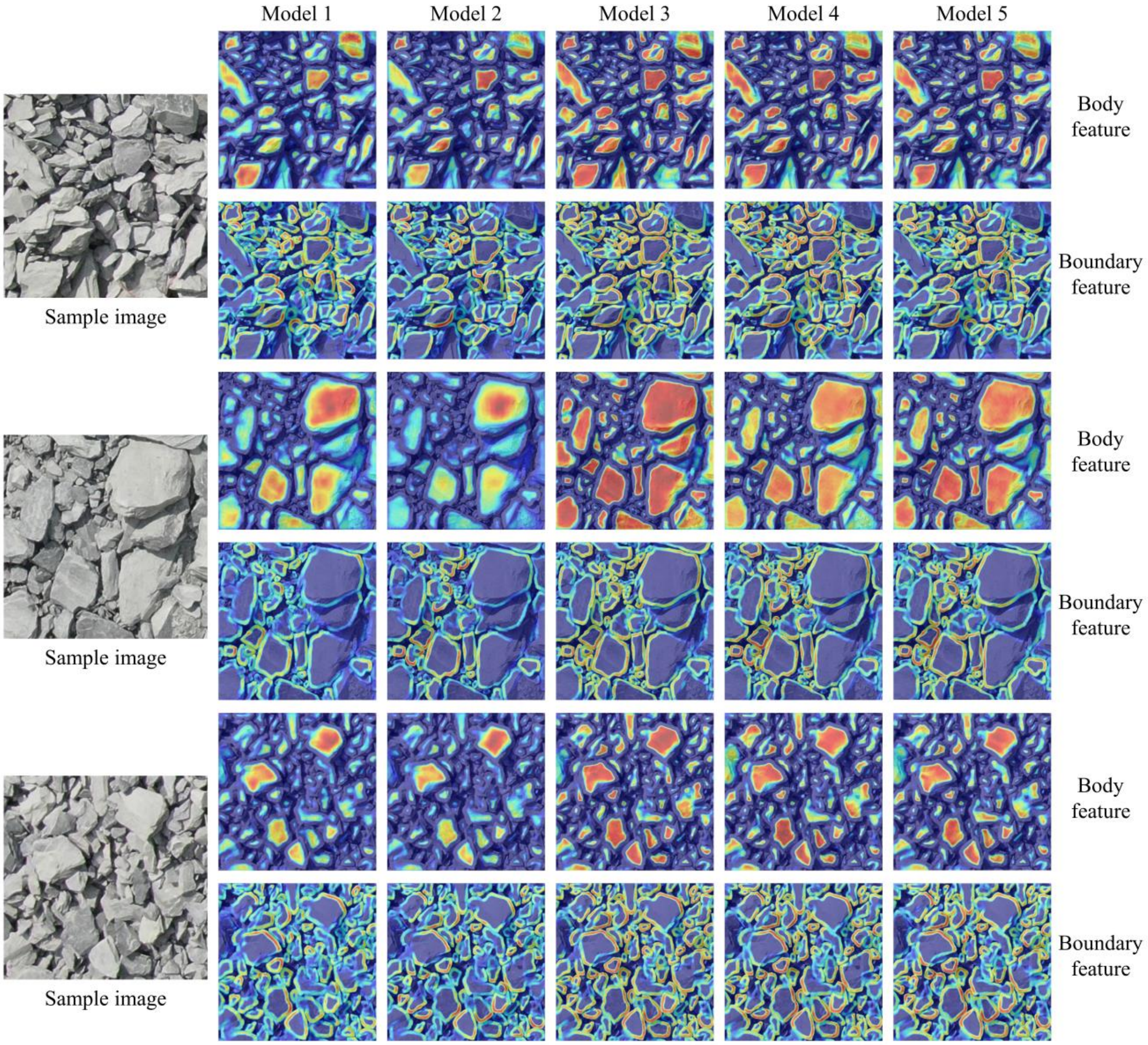

***Fig. 11.*** *Visualization of body and boundary feature maps for five models, with regions of high responsiveness highlighted in red.*

### *3.1.3 Patch stitching and post-processing*

The training and evaluation of the improved Unet semantic segmentation model were based on 512 × 512 image patches, while our objective is to process images of full size. Additionally, although our semantic segmentation model identifies the body and boundary of particles, it does not segment each particle completely. To address these issues, we employed stitching predictions and post-processing techniques.

To predict a new full-size image, the entire image is divided into a series of 512 × 512 image patches using a sliding window approach. These patches are individually processed by the improved Unet semantic segmentation model, with the outcomes subsequently stitched together. During prediction, morphological operations employing the opening operator with a 9 × 9 structuring element are used to eliminate small regions that constitute noise. Prior to the morphological operation, body and boundary pixels are assigned the same grayscale value, which is then reverted back to their original grayscale values after morphological processing. This strategy of combining the body and boundary pixels before morphological operations helps prevent small body pixels enclosed by boundary pixels from being erroneously removed, thus ensuring no loss of potential region seed points in the post-processing phase.

Using the sliding-window-based stitching prediction and the post-processing method described in Section 2.3.3, we can obtain the final rock particle segmentation results. Fig. 12 shows some typical segmentation predictions, where each unique color block represents a separate particle. The segmentation results for the majority of the particles in the images are sufficiently accurate, despite a few imprecise instances. The main issue with these inaccuracies relates to some particles not being segmented apart, resulting in a segmented block that is actually a composite of two or more particles. This phenomenon is likely due to the semantic segmentation model's imperfect extraction of boundary pixels; when two or more particles are closely adjacent, the boundary pixels at their juncture are not completely retrieved, allowing the body pixels of two or more particles to remain connected. During post-processing, this connected region is treated as a single seed point, leading to a combined particle in the final outcome. Similarly to other particle segmentation studies, this study encounters challenges in accurately segmenting fine particles with pixel diameters smaller than 8 to 10 pixels, which corresponds to an actual size of about 1 to 1.25 centimeters. These discrepancies contribute to certain inaccuracies in the calculated particle size distribution results.

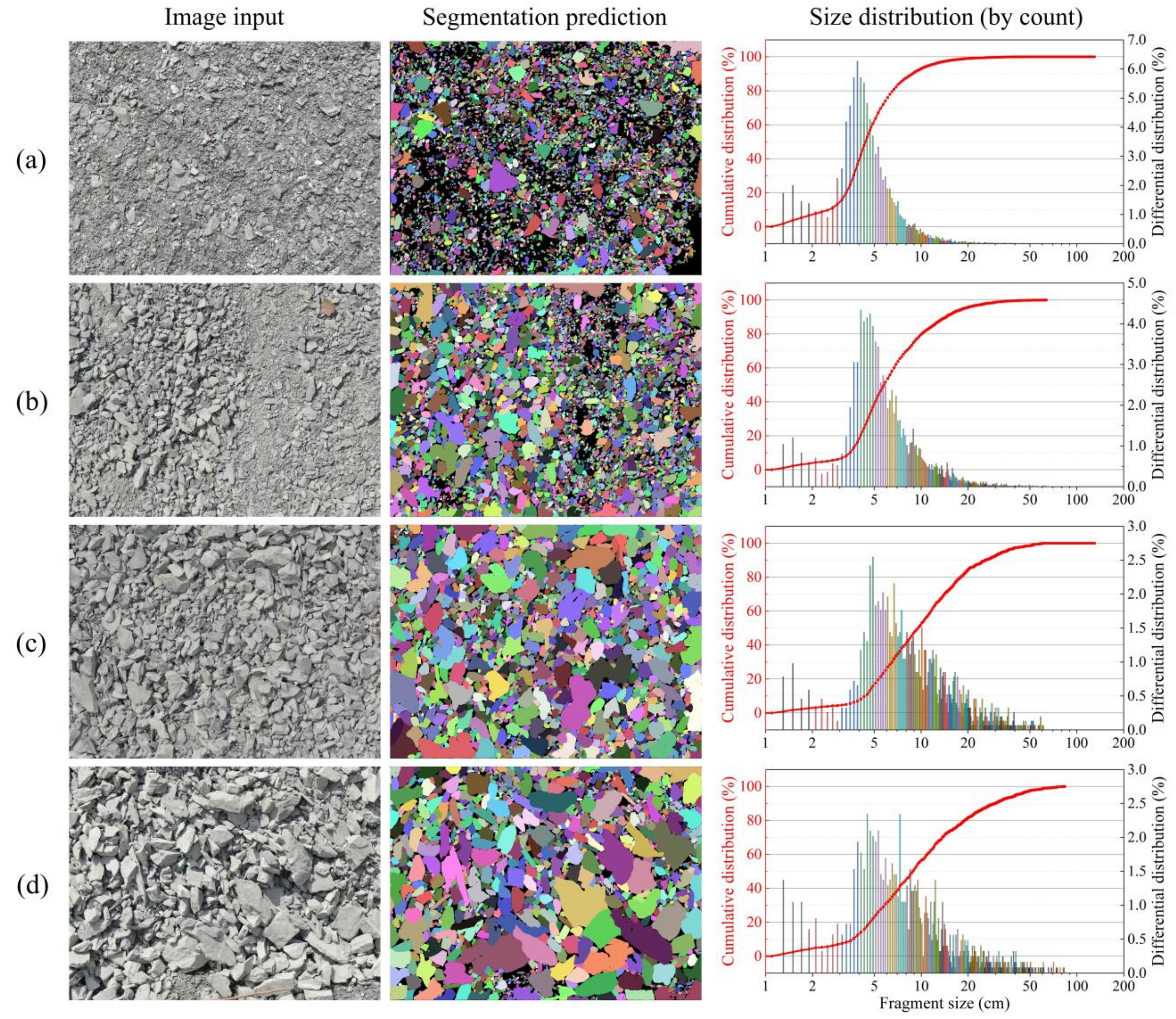

***Fig. 12.*** *Final segmentation results from four representative images selected respectively from Section 1, 2, 3, and 4.*

### ***3.2 Fragment size distribution***

#### *3.2.1 Size distribution by count*

Size analysis of rock fragments involves interest in size distribution by count. For each particle segmented, its shape is approximated by an equivalent ellipse, and the equivalent ellipse diameter *d* is calculated to represent the fragment size. This size is then converted from pixels to actual measurements in centimeters based on image calibration results. In the far-right column of Fig. 12, size distribution (by count) corresponding to each segmentation prediction is displayed. Considering the large number of particles in each image (exceeding 6000 in some cases), histogram bins were constructed with a differential interval of 0.2 cm to prevent overcrowding. Since the x-axis is logarithmic, some spacing or overlap exists between bins. The cumulative distribution curve of particle counts is depicted by the red dotted line in the charts. It should be noted that fine particles with a pixel diameter of 10 pixels or less were not segmented satisfactorily. Therefore, these particles were excluded from the statistical analysis.

Combining segmentation prediction results with their respective size distribution charts allows for a clear visualization of the distinct variations in rock fragment size distribution across different Sections. Qualitatively, Section 1 (Fig. 12 (a)) has the smallest overall particle size with a dense presence of small-sized fragments, as reflected by the leftward skew of histogram bins and only a few bins in the larger size ranges. The cumulative distribution curve in this section is steeper, which suggests that the accumulation of total particle count mainly occurs within smaller size ranges. Section 4 (Fig. 12 (d)), however, exhibits the largest overall particle size, with more densely populated bins in the larger size ranges, indicating that a larger portion of the cumulative total particle count is attributed to larger-sized particles. It is also noticeable that the histogram in Section 1 shows significant peaks and high variability, while the histogram for Section 4 is flatter with less variation. Quantitatively, in Fig. 12 (a), 61.96% of particles are smaller than 5cm, with only 1.00% being larger than 20 cm, and the average particle size is 5.39 cm. In Fig. 12 (b), 39.02% of particles are smaller than 5 cm, with 4.39% being larger than 20 cm, yielding an average particle size of 7.81cm. For Fig. 12 (c), particles under 5cm account for 16.74% of the total, while those exceeding 20 cm comprise 15.64%, with an average size of 12.79cm. Lastly, Fig. 12 (d) shows that 23.10% of particles are smaller than 5 cm, with a significant 18.23% larger than 20 cm, leading to an average size of 13.33 cm. The higher proportion of particles under 5 cm in Fig. 12 (d) compared to Fig. 12 (c) is due to the fact that Fig. 12 (d) actually includes many fine particles despite having a generally larger overall particle size. In summary, from Section 1 to Section 4, the average particle size increases progressively, transitioning from dominantly fine particles to coarser ones.

*3.2.2 Size distribution by volume*

The number distribution results for rock fragments in the four divisions already highlight the spatial heterogeneity of the pile's particle size distribution. However, considering practical engineering needs, the handling and utilization of stone material are often measured by mass (or volume, under the constant density assumption), making the analysis of size distribution by volume equally important.

Once particle segmentation is complete, each rock fragment is equated to an ellipsoid, and its volume is represented by this geometrical figure, with the fragment size still determined by the equivalent ellipse diameter $d$. Using a methodology similar to that employed in Section 3.2.1, histograms and cumulative distribution charts for size distribution by volume are generated, as shown in Fig. 13. These sub-figures correspond to the ones in Fig. 12, with the height of each histogram bar representing the total volume percentage of particles within a specific size differential interval set at 0.8 cm for optimal visual presentation. Additionally, characteristic particle diameters $d_{10}$, $d_{50}$, $d_{90}$ (i.e., the particle diameters corresponding to 10%, 50% and 90% volume passing, respectively) are marked in the graphs with their respective positions and values.

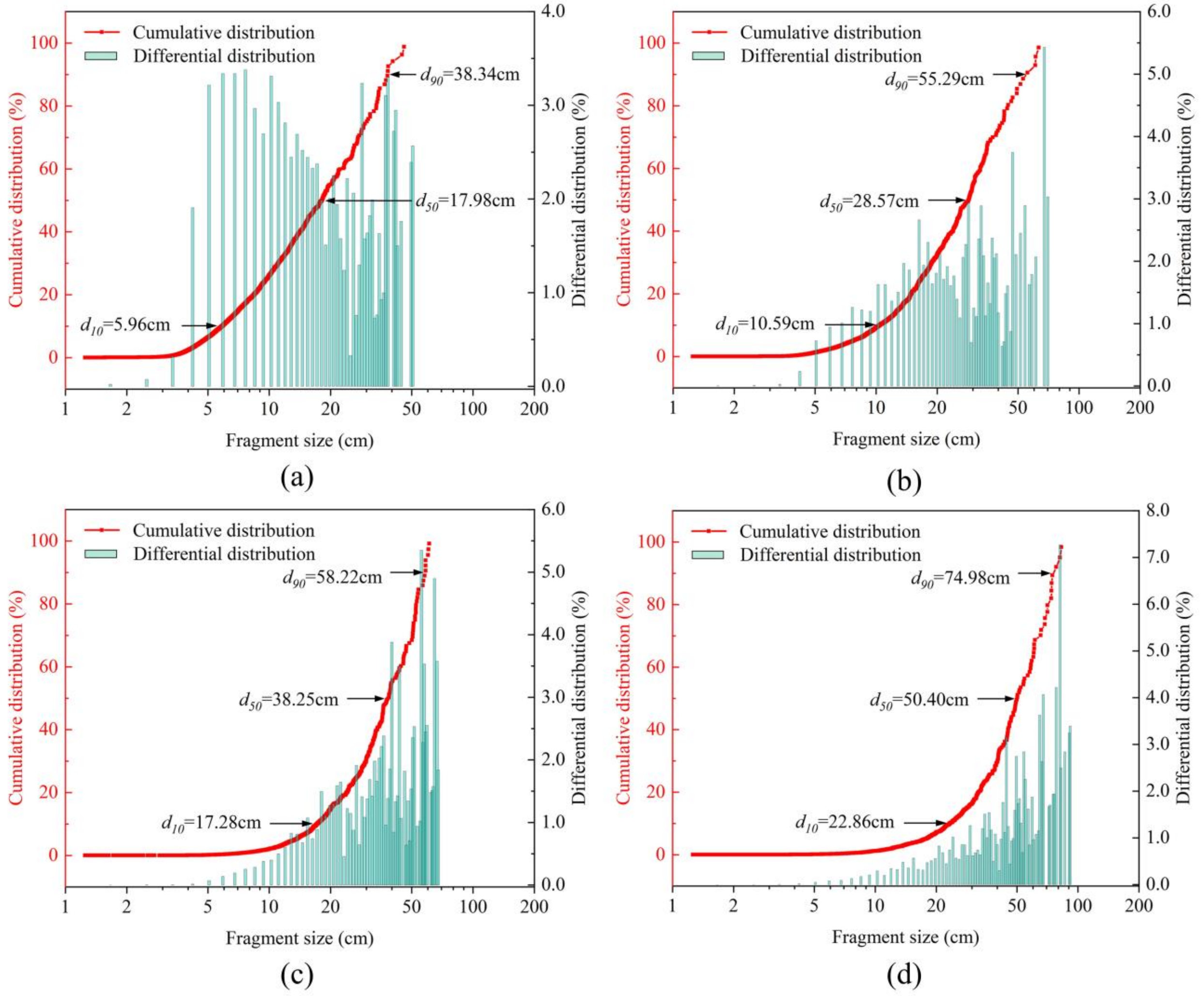

***Fig. 13.*** *The size distribution of rock fragments (by volume).*

Across the sub-figures in Fig. 13, the notable differences in particle distribution between Sections can still be observed. Specifically, histograms transition from a flat, low-variance distribution in Section 1 to a right-skewed distribution with distinctive peaks in Section 4. Cumulative curves likewise show a gradual shift from a gentle ascent in Section 1 to a steeper rise on the right side in Section 4, indicating a progressive concentration of data values in the higher ranges, i.e., an increase in the volume proportion of larger particles. On a quantitative note, for Section 1, the smallest characteristic diameter $d_{10}$ is 5.69 cm, whilst for Sections 2, 3, and 4 this value rises to 10.59 cm, 17.28 cm, and 22.86 cm, respectively, meaning that particle volume accumulation at the smallest diameters continues up to 22.86 cm in Section 4 before reaching 10% of the total volume. The median size $d_{50}$ increases from 17.98 cm to 28.57 cm, 38.25 cm, and 50.40 cm through the Sections. A similar upward trend is observed for d90, with Section 4's $d_{90}$ at 74.98 cm, nearly double that of Section 1. Overall, the distribution charts by volume also reflect a shift from being dominated by smaller particles to larger ones from Section 1 to Section 4. Differing from the count-based statistics, volume-based distribution charts emphasize the high volume proportion of large particles more.

### *3.3 Spatial segregation of fine and corse particles*

In Section 3.2's fragment size distribution analysis, representative images from Sections 1 to 4 were selected for particle size analysis. This analysis highlighted differences in size distribution between sections, revealing a pattern of finer particles at the top of the slope and coarser particles at the bottom, thus preliminarily indicating a segregation phenomenon along the slope.

Further granularity distribution analysis of 5 non-overlapping images from each section, totaling 20 images, was conducted. The average values and 95% confidence intervals (CI) for the characteristic particle diameters ($d_{10}$, $d_{50}$, $d_{90}$) of each section were computed and are listed in Table 4. All characteristic diameters show an increasing trend with depth, indicating overall particle segregation in the vertical direction, with fine particles tending to concentrate at the top of the slope while coarser particles distribute towards the bottom. Within specific depth ranges, characteristic diameters vary, showing some lateral heterogeneity in size distribution. Notably, the variability range of $d_{10}$ is much smaller than that of $d_{50}$ and $d_{90}$. The size of the variability range generally increases with depth, with some overlap evident in Section 4 between the 95% CI intervals of $d_{50}$ (40.50-75.99) and $d_{90}$ (60.66-117.40). Fig. 14 visualizes the results from Table 4.

These findings raise an interesting question regarding which characteristic diameter is more effective in representing particle size on the slope. Analysis indicates that $d_{50}$ and $d_{90}$ are more susceptible to the influence of extremely large particles, hence displaying higher variability and uncertainty. This is why there are overlapping intervals of characteristic diameters between consecutive sections. Conversely, due to the steady change and consistency of $d_{10}$, it is suggested as a more effective measure for particle size representation.

*Table 4* *Mean percentile values (d10, d50, d90) and 95% confidence interval (CI)*

| Section | Depth | $d_{10}$ (cm) | 95% CI (cm) | $d_{50}$ (cm) | 95% CI (cm) | $d_{90}$ (cm) | 95% CI (cm) |
|---|---|---|---|---|---|---|---|
| 1 | 0~12m | 5.77 | [4.42-6.69] | 18.01 | [12.11-22.14] | 37.55 | [29.10-44.76] |
| 2 | 12~24m | 8.30 | [7.13-9.21] | 24.71 | [20.92-28.72] | 45.13 | [39.55-50.27] |
| 3 | 24~36m | 13.11 | [10.48-17.28] | 35.38 | [30.12-42.25] | 59.14 | [47.97-74.88] |
| 4 | 36~48m | 27.63 | [21.28-32.51] | 60.89 | [40.50-75.99] | 97.64 | [60.66-117.40] |

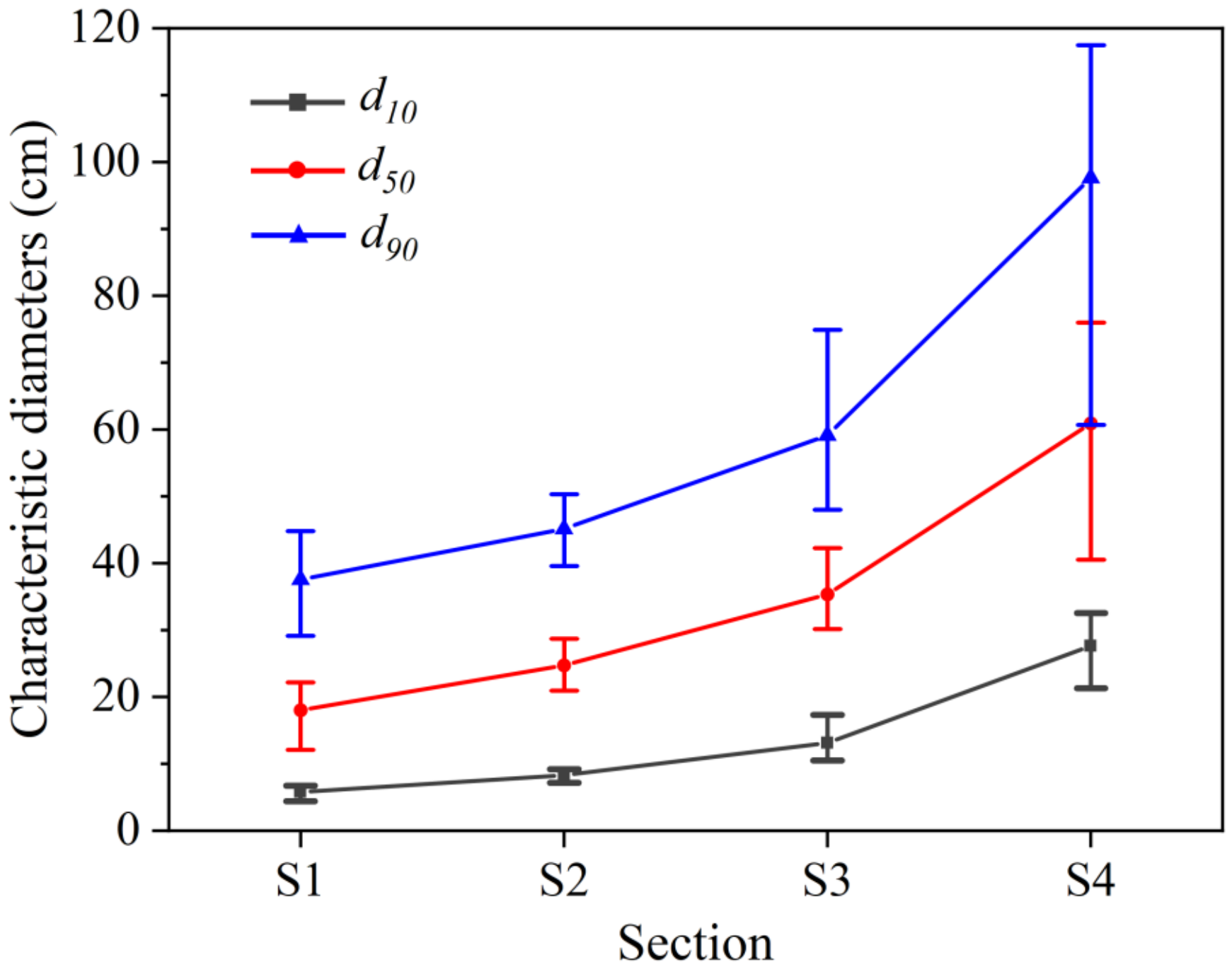

***Fig. 14.*** *Characteristic diameters (d10, d50, d90) with 95% confidence interval from all images for a particular section.*

Compiling all particles from the 20 images to draw a particle size distribution curve, as shown in Fig. 15, represents the overall particle size distribution of the slope. The values of $d'_{10}$, $d'_{50}$, $d'_{90}$ are marked as 11.08 cm, 37.72 cm, and 76.70 cm, respectively. Calculating relative characteristic diameters $d_{10}/d'_{10}$ ($d_{50}/d'_{50}$, $d_{90}/d'_{90}$) further deepens the analysis of the segregation phenomenon. The average value of $d_{10}/ d'_{10}$ increased from 0.52 in Section 1 to 2.49 in Section 4, marking a 379% increase. The average value of $d_{50}/ d'_{50}$ increased from 0.47 to 1.61, a 243% increase. A similar trend is observed for $d_{90}/ d'_{90}$, which increased from 0.49 to 1.27, marking a 159% increase. Relative characteristic diameters in Sections S1 and S2 are generally less than 1, in Section S4 generally greater than 1, while in Section S3 they can be either above or below 1. As illustrated in Fig. 16, the relative characteristic diameters for each section can be approximated by a straight line fit. The increase rate of relative characteristic diameters decreases from 0.636 for $d_{10}/ d'_{10}$ to 0.253 for $d_{90}/ d'_{90}$, indicating that smaller particles are more sensitive to segregation.

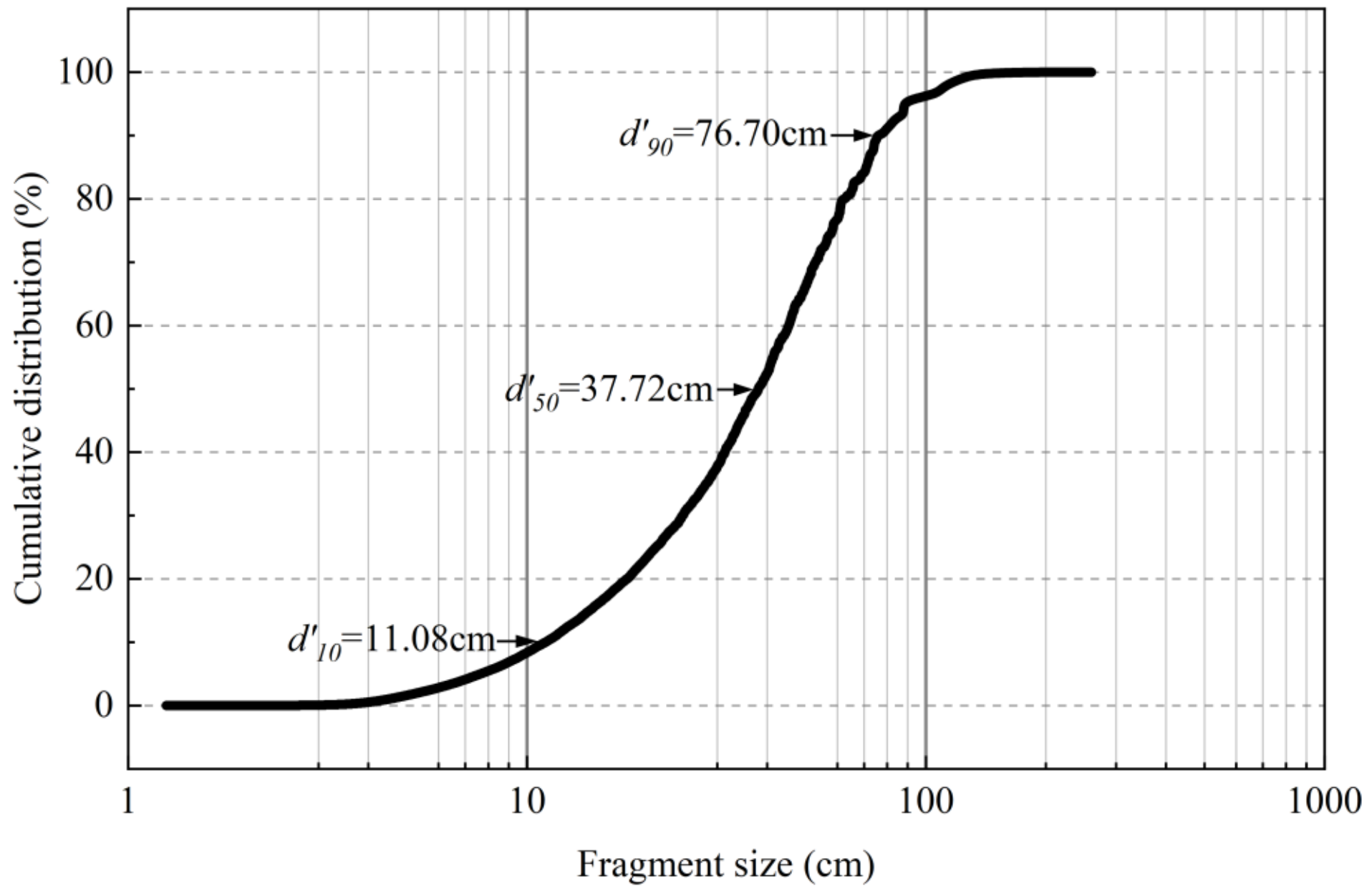

***Fig. 15.*** *Size distribution curve of the whole slope.*

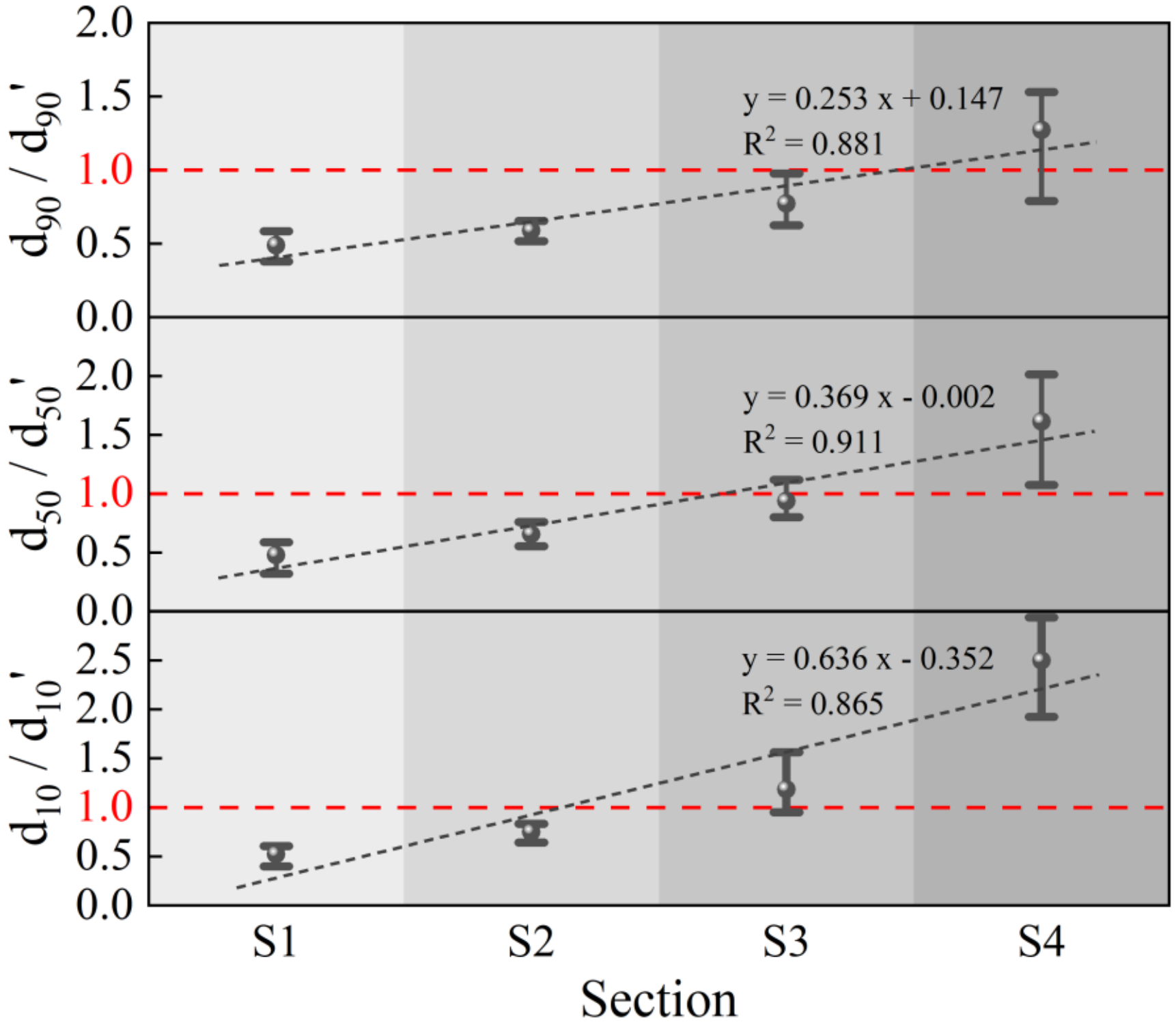

***Fig. 16.*** *Relative characteristic diameters.*

## 4. Conclusions

This study successfully mapped the segregation pattern of rock fragments in a quarry through an innovative application of deep learning techniques in aerial image analysis. A series of twenty aerial images captured by drones underwent detailed analysis, employing an improved Unet semantic segmentation model complemented by a post-processing strategy involving region expansion. An ellipsoid model was employed to approximate the irregular shapes of the fragments. A vertical partitioning of the slope into four distinct sections allowed for the analysis and comparison of fragment size distributions, critically demonstrating the spatial differentiation of particle sizes within these sections. The primary findings are summarized as follows:

(1) The enhanced semantic segmentation model demonstrated excellent performance in segmenting particles, accurately extracting the bodies and boundaries of the fragments. The introduced CARAFE, GhostConv and ECA modules each contributed to improvements in model performance. Following the semantic segmentation model's extraction of bodies and boundaries, the proposed region expansion post-processing algorithm enabled the accurate segmentation of complete particles.

(2) There was a discernible variance in the morphological characteristics of the size distribution histograms and curves among the different depth sections, denoting vertical heterogeneity across the slope. The comparative visualization of size distributions, based on both count and volume, highlights these differences, notably endorsing the use of volume-based charts for a more robust interpretation and suggesting the derivation of characteristic particle sizes from cumulative distribution analyses.

(3) A manifest pattern of segregation was observed along the slope's vertical gradient, as evidenced by the stepwise rise in the values of relative characteristic diameters from the slope's crest to its toe. This trend illustrated a clear segregation, with finer particles disproportionately located towards the top of the slope and relative characteristic diameters hovering near the value of 0.5, in stark contrast to the base of the slope where coarser particles aggregated, increasing the relative characteristic diameters by an impressive rate of 159% to 379%.

The findings of this study highlight both the heterogeneity of fragment size distribution in quarry materials and the effectiveness of deep learning and image analysis in elucidating complex geotechnical phenomena. The insights present significant ramifications for optimizing the storage, transportation, and allocation of quarry material.

## Acknowledgements

The financial supports of the National Natural Science Foundation of China (Grant Nos. 52009036, 51979091) and Key Technologies R & D Program of Yunnan Province, China (Grant No. 202202AF080003) are greatly acknowledged.